# Quantifying Flat-Band Voltage in Si Metal-Oxide-Semiconductor Structures: An Evaluation via Terahertz Emission Spectroscopy (TES)

Dongxun Yang, Tonouchi Masayoshi [a]

Institute of Laser Engineering, Osaka University, 2-6 Yamadaoka, Suita, Osaka 565-0871, Japan

[a] tonouchi.masayoshi.ile@osaka-u.ac.jp

## Abstract

Laser-induced Terahertz (THz) Emission Spectroscopy (TES) has demonstrated its potential utility in the realm of Metal-Oxide-Semiconductor (MOS) devices as an expedient and noncontact estimation methodology. Owing to its discerning response to the interface electric field, the amplitude of the THz emission peak in time-domain spectroscopy encapsulates rich information regarding MOS properties, notably the flat-band voltage. This paper concentrates on the precise quantitative estimation of the flat-band voltage within the Si MOS structure, elucidating the intricacies of the estimation process through the THz emission model.

## I. INTRODUCTION

The Si Metal-Oxide-Semiconductor (MOS) structure constitutes a fundamental element in contemporary semiconductor technology[1]. The expeditious and judicious estimation of pertinent parameters within this structure, notably the interface potential and flat-band voltages, is imperative[2,3]. Terahertz Emission Spectroscopy (TES) emerges as a potential noncontact methodology, facilitating the rapid and semi-quantitative characterization of various semiconductor structures[4] including Si MOS structures, with a specific emphasis on the expeditious estimation of critical parameters.[5] In this paper, an innovative approach employs the intrinsic correlation between the amplitude of terahertz (THz) emissions and the externally applied DC bias to the Si MOS structure, and the precise quantitative estimation of the flat-band voltage within the Si MOS structure is elucidated. The proposed method underscores its utility in the swift assessment and discernment of semiconductor characteristics, presenting a promising avenue for efficiency and efficacy in the semiconductor industry.

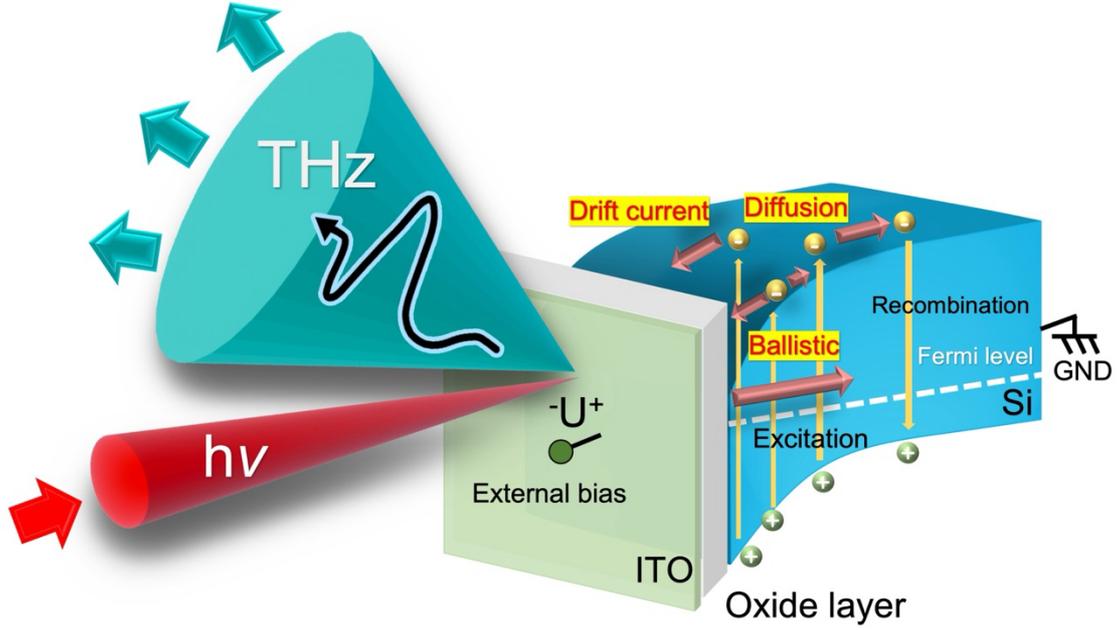

**Figure 1** Mechanism of THz emission from Si MOS structures with external bias.

Ultrafast charges transport leads to the THz radiation[6-12]. With an ultrafast laser pulse illuminating the semiconductor to generate the photocarriers, the movement of the transient photocarriers radiates the THz wave. According to the types of ultrafast charge transport, the mechanism of the THz emission can be mainly concluded as 1) carrier drift current accelerated by electric field[13, 14]; 2) diffusion current by density gradient[15, 16]; 3) carrier diffusion by quasi-ballistic hot electron motion[17]. A diagram of the THz emission mechanism from Si MOS structures with external bias is shown in Fig.1. The drift current relies on the association between the electric field, carrier mobility, and optical penetration depth. During the fast transient period of approximately 200-300 fs, fs laser illumination induces a photocurrent that is directly proportional to the laser intensity. The THz emission field can be simply derived as,

$$E_{THz} \propto \frac{dJ}{dt} \propto \frac{\Delta n \Delta v}{\Delta t} \propto \mu E_B I_p, \qquad (1)$$

where $E_B$ is the built-in electric field, $I_p$ is the intensity of the laser, and $\mu$ is the carrier mobility. The diffusion current includes the part by the density gradient or the photo-Dember effect and the part by the quasi-ballistic hot electron motion. When the energy of the hot photocarriers is significantly higher than the conduction bands, it leads to hot photocarrier injection. In such a scenario, the THz emission mechanism occurs due to quasi-ballistic transport of high-energy carriers and the THz emission field can be replaced by,

$$E_{THz} \propto I_p \sqrt{\frac{E_p - E_g}{m^*}} \tag{2}$$

where $E_p$ and $E_g$ represent the photon energy and the bandgap of the semiconductor. $m^*$ is the effective mass of the electrons. Therefore, the THz emission due to the diffusion current is always constant when the laser power and samples have been decided.

## II. THz emission theory model

Here we provide the simplified THz emission model of MOS structure which elucidates the relationship between THz emission amplitude and external bias applied to Si MOS structures. Generally, the basic formula of THz emission from the semiconductors has been given in Eq. (2), in which the built-in field in the Si MOS structure can be simply obtained with depletion approximate assumption. Therefore, the THz emission field can be calculated from interface potential ($V_s$) as follows,

$$E_{THz} \propto \frac{dJ}{dt} \propto \mu \frac{V_s}{w} I_p \propto \mu \sqrt{\frac{N_i V_s}{\varepsilon_0 \varepsilon_r}} I_p . \tag{3}$$

where $w$ is the thickness of depletion layer, $N_i$ is the impurity density of Si, $\varepsilon_0$ and $\varepsilon_r$ is the dielectric constant of the vacuum and Si.

In the ideal Si MOS structure, the external bias declines from the gate to the Si bulk where the electric potential is zero. Therefore, the external bias is equal to the sum of interface potential and potential drop across the oxide layer, expressed as

$$V_g = V_s + V_{ox} \tag{4}$$

where $V_g$ is the external bias, $V_s$ and $V_{ox}$ are the interface potential and potential drop across the oxide layer. In addition, the capacitance of the oxide layer and the surface of the semiconductor are series connected; therefore, the relationship between $V_{ox}$ and $V_s$ is expressed as

$$\frac{V_s}{V_{ox}} = \frac{C_{ox}}{C_s}, \tag{5}$$

where $C_{ox} = \varepsilon_0 \varepsilon_{ox}/d$ is the capacitance of the oxide layer in a unit area. Based on these relations, $V_g$ and $V_s$ can be correlated as follows:

$$V_g = V_s \left(1 + \frac{C_s(V_s)}{C_{ox}}\right) \tag{6}$$

The Eq. (3) and Eq. (6) constitute the THz emission theory model of Si MOS, from which the relationship between the THz emission intensity and external bias voltages has been clarified[5]. In return, the parameters, including the interface potential and flat-band voltage of the Si MOS structure can be estimated.

## III. Experiment and Discussion

In this part, our focus revolves around the assessment of the Si MOS through the comparison of the experimental $A{\sim}V_g$ curve with the calculated $E_{THz}{\sim}V_g$ curve, where the parameter $A$ represents the amplitude of the first peak in the THz emission time-domain spectrum and $E_{THz}$ represents the calculated THz emission field based on the THz emission theory model. The experiment has been carried out in both n-type and p-type Si MOS structures under an 800-nm laser pulse illumination with external bias ranging from -10 V to 10 V. The samples were excited at 80 MHz using an 800-nm source at an incident angle of 45°, and the THz emission was focused onto the GaAs THz sensor with variable delays. The average excitation power was set to 100 mW. The details of the experimental results can be found in Ref. [13]. The calculated parameters of the laser, oxide layer, and Si are listed in the Tables. 1, 2, and 3, respectively[13].

**Table 1** Parameters of laser used in this study.

| $\lambda$(nm) | $r$(mm) | $P$(mW) | $I_p$(W/m$^2$) |
|---|---|---|---|
| 800 | 5 | 100 | 2546.55 |

$\lambda$ = laser wavelength; $r$ = radius of laser beam; $P$ = laser power; $I_p$ = laser intensity.

**Table 2** Parameters of oxide layer.

|  | $\varepsilon_{ox}$ | $d$(nm) | $\varepsilon_0$(F/m) | $C_{ox}$(mF/m$^2$) |
|---|---|---|---|---|
| p-type | 3.9 | 120 | 8.854E-12 | 0.286 |
| n-type | 3.82 | 90 | 8.854E-12 | 0.374 |

$\varepsilon_{ox}$ = dielectric constant of oxide layer; $d$ = thickness of oxide layer; $\varepsilon_0$ = vacuum dielectric constant; $C_{ox}$ = capacitance per square meter of oxide layer.

**Table 3** Parameters of p- and n-type Si samples.

| $N_A$(cm⁻³) | $N_D$(cm⁻³) | $n_i$(cm⁻³) | $\varepsilon_r$ | $k$(eV/K) | $E_g$(eV) | $T$(K) | $V_{FB,p}$(V) | $V_{FB,n}$(V) |
|---|---|---|---|---|---|---|---|---|
| 5.5E15 | 9.3E14 | 9.7E9 | 12 | 8.62E-5 | 1.12 | 300 | -1.60 | -0.96 |

$N_A$ = doping density of acceptors for p-type Si; $N_D$ = doping density of donors for n-type Si; $n_i$ = intrinsic carrier density; $\varepsilon_r$ = dielectric constant of Si; $k$ = Boltzmann constant; $E_g$ = bandgap energy of Si; $T$ = room temperature; $V_{FB,p}$ and $V_{FB,p}$ = flat-band voltages of the p- and n-type Si MOS samples, respectively.

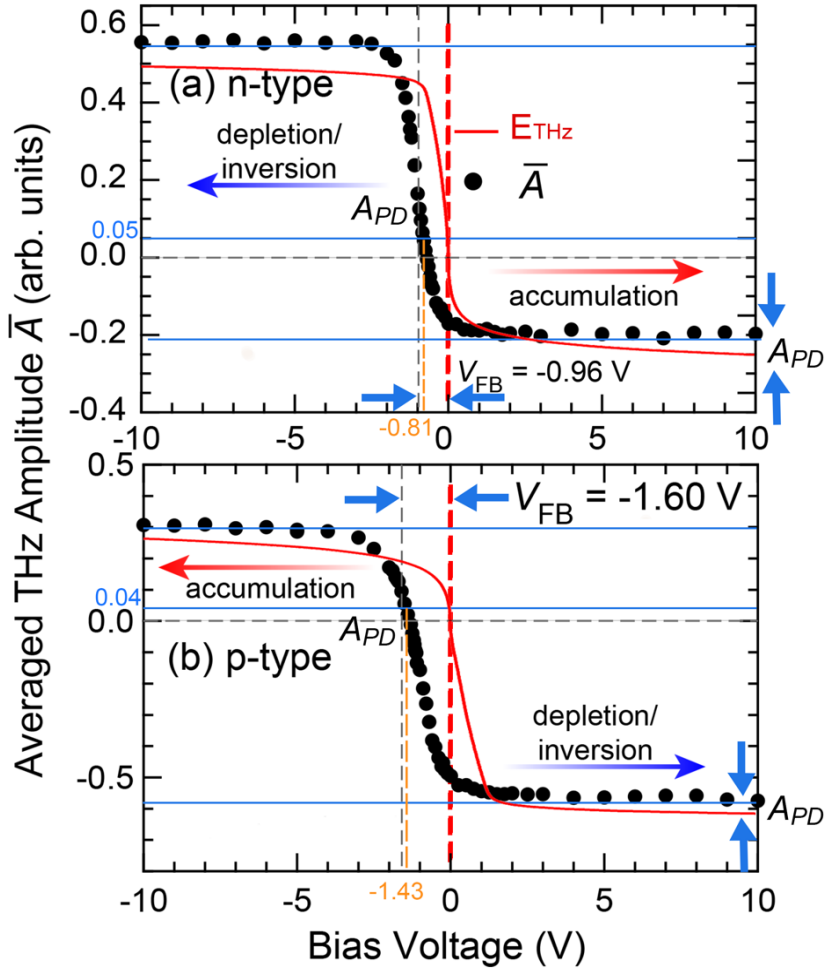

**Figure 2** The calculation results of the $E_{THz} - V_g$ curves in (a) n-type Si MOS and (b) p-type Si MOS and their relative $E_{THz} - V_g$ curves. The original experiment data are taken from Ref. [12].

Figure 2 shows the comparison between the calculated $E_{THz} \sim V_g$ curve by using the model under ideal conditions and the experimental $A \sim V_g$ data for both the n- and p-type Si MOS samples[13]. Based on the comparison in Fig.2, significant disparities between calculated results and experimental observations are apparent along both the horizontal and vertical axes. These disparities are primarily attributed to the flat-band voltage and the photo-Dember effect. Within our model, we posit an ideal condition wherein the flat-band voltage is assumed to be zero, and THz emission solely arises from the drift current. However, in practical Si MOS structures, the inherent work-function distinction between the gate material (ITO in our case) and the Si substrate induces initial interface band bending, thereby expediting the movement of laser-excited photocarriers and leading to THz wave radiation without external bias.

In our experimental setting, a noticeable THz emission signal is detected from the Si MOS even when the external bias is absent. This occurrence is explicable by the work-function difference inducing initial band bending, facilitating THz emission without the need for external bias. Introducing an equivalent reverse voltage to the Si MOS to rectify the initial band bending to a flat state designates this voltage as the flat-band voltage. Consequently, the THz emission intensity diminishes significantly, though not to zero, as the drift current becomes null, ceasing its contribution to THz emission. The deviation observed along the horizontal axis is thus elucidated by the presence of initial band bending, and conversely, the flat band voltage can be deduced from this deviation.

Another source of deviation along the vertical axis arises from THz emission resulting from the diffusion current due to the photo-Dember effect[15]. Laser illumination induces the generation of photocarriers in the semiconductor, and THz emission tied to the photo-Dember effect persists across various external bias voltages, including the flat band voltage condition. The intensity of this THz emission, contingent upon the bandgap of the semiconductor, remains constant. Hence, a rapid estimation of parameters can be facilely achieved by comparing the deviations between experimental and calculated results.

Here we introduce the detailed steps of the parameter estimation. First, standardization of $E_{THz}$ and $A$. Since the value of calculated $E_{THz}$ and experimental $A$ are arbitrary units, we need to confirm the horizontal range are same for both calculation and experimental data, which means the absolute value of the difference between maximum value and minimum value should be standard and equivalent for both $E_{THz}$ and $A$ to make them comparable. Second, estimation of the THz emission from diffusion current ($A_D$). Roughly, we can directly obtain the value of $A_D$ by putting the calculated $E_{THz}$ and experimental $A$ together into one chart and find out the difference between $E_{THz}$ and $A$. We can also estimate the $A_D$ by using the following formula as,

$$A_D = \frac{\sum_{i=1}^{n}(A_i - E_{THz,i})}{n} \tag{7}$$

The average of the difference between $E_{THz}$ and $A$ can be regarded as the value of the $A_D$. When $n$ becomes larger, the value of $A_D$ becomes more accurate. Third, estimation of flat band voltage. At the flat band voltage, only the diffusion current takes charge of the THz emission. Therefore, we can easily obtain the flat band voltage from the $A - V_g$ curve when $A = A_D$,

$$V_{FB} = V_g(A = A_D) \tag{8}$$

Table.4 The estimation results of $A_D$ and $V_{FB}$ for both n- and p-type Si MOS.

|  | $A_{D,n}$ (a.u.) | $A_{D,p}$ (a.u.) | $V_{FB,n}$ | $V_{FB,n}^*$ | $V_{FB,p}$ | $V_{FB,p}^*$ |
|---|---|---|---|---|---|---|
| Value | 0.05 | 0.04 | -0.81 V | -0.96 V | -1.43 V | -1.60 V |

*The flat-band voltage estimated from C-V measurement.

Table 4 delineates the flat-band voltage estimates derived from $A - V_g$ curve with the combination of the THz emission theory model, alongside corresponding results obtained through C-V measurements. This approach enables a quantitative assessment of the flat-band voltage in Si MOS structures solely through TES, eliminating the necessity for supplementary methods like C-V measurements. This streamlined methodology enhances the practical utility of TES. Moreover, leveraging the calculated outcomes from the THz emission model, the quantitative association between $E_{THz}$ and $V_g$ offers the capacity to estimate additional parameters, including $V_s$ and the various factors such as doping concentration of Si, dielectric constant and thickness of oxide layer, work-function of gate material[5, 18], etc.. This extends the quantitative characterization capabilities of TES beyond the flat-band voltage, thereby providing a comprehensive insight into the intricate parameters governing Si MOS structures.

## IV. Conclusion

In conclusion, our investigation into Si MOS structures through TES has yielded valuable insights and quantitative estimations. By analyzing the experimental $A - V_g$ curve and comparing it with the calculated $E_{THz} - V_g$ curve under ideal conditions, we successfully quantified the flat-band voltage for both n- and p-type Si MOS samples. Remarkably, our approach demonstrates the efficacy of TES as a standalone, noncontact method for flat-band voltage estimation, obviating the need for complementary techniques like C-V measurements. This methodology not only advances our understanding of fundamental Si MOS properties but also highlights the practical utility of TES in the semiconductor industry for rapid and accurate parameter estimation.

## Acknowledgments

M.T. acknowledges support in part by JSPS KAKENHI Grant No. JP 23H00184, and JST, CREST Grant Number JPMJCR22O2, Japan. The authors acknowledge Mochizuki T. *et.al.* for providing experiment data and information.